\documentclass[prb,twocolumn,superscriptaddress,amsmath,amssymb,showpacs,floatfix,preprintnumbers]{revtex4}
\usepackage{amsmath}
\usepackage{graphicx}
\usepackage{dcolumn}
\usepackage{color}
\usepackage{lipsum}
\DeclareGraphicsExtensions{.png .jpg .pdf}

\begin{document}

\title{Dirac spectrum in gated multilayer black phosphorus nanoribbons}

\author{J. D. S. Forte}\email{johnathas@fisica.ufc.br}
\affiliation{Departamento de F\'isica, Universidade Federal do Cear\'a, Caixa Postal 6030, Campus do Pici, 60455-900 Fortaleza, Cear\'a, Brazil}
\author{D. J. P. de Sousa}\email{duarte.j@fisica.ufc.br}
\affiliation{Departamento de F\'isica, Universidade Federal do Cear\'a, Caixa Postal 6030, Campus do Pici, 60455-900 Fortaleza, Cear\'a, Brazil}
\author{J. Milton Pereira Jr.}\email{pereira@fisica.ufc.br}
\affiliation{Departamento de F\'isica, Universidade Federal do Cear\'a, Caixa Postal 6030, Campus do Pici, 60455-900 Fortaleza, Cear\'a, Brazil}

\date{ \today }

\begin{abstract}
We investigate the effects of a perpendicular electric field applied to multilayer phosphorene nanoribbons with zigzag and armchair edges. Within the context of the tight-binding model, we explore the electronic properties of these systems giving emphasis to the appearance of Dirac-like spectra, a transition that occurs when the gate density associated with the applied displacement field is greater than the critical value $n_c$. We show that the confinement properties and the screening effects in such systems play an important role on the determination of $n_c$, suggesting a scheme to determine the thickness, width and edge orientation of multilayered phosphorene nanoribbons. We also explore how this transition affects the electronic transport properties of such systems.
\end{abstract}

\pacs{71.10.Pm, 73.22.-f, 73.63.-b}
\maketitle

\section{Introduction}
Phosphorene, a single layer of black Phosphorus (BP), is a relatively new 2D semiconductor which has gained a lot of attention since its production in 2014\cite{ref1}. Due to its interesting properties, such as high carrier mobility\cite{ref1} and thickness-dependent energy gap\cite{ref2, ref3, ref4}, multilayer BP is considered a promising material with a great potential for applications in nanoelectronics and optoelectronics\cite{ref5,ref6, ref7,ref8, ref9}. In particular, the wide range of values that its thickness-dependent energy gap can assume, ranging from $\approx 2.0$ eV for the monolayer to $\approx 0.3$ eV for the bulk BP, has significant importance since it covers a broad range of applications not reached by other 2D semiconductors, such as in fiber optic telecommunication and thermal imaging\cite{ref10}.

In addition to the thickness variation, the energy gap of multilayer phosphorene systems can also be tuned through the application of an external perpendicular electric field\cite{ref11, ref12, ref13, LLL, ref15, ref16}. It was both theoretically predicted\cite{ref12, ref13} and experimentally observed\cite{ref17, ref18} that such mechanism induces topological phase transitions for fields at the vicinity of a critical value $D_c$ for which a gap closure is achieved. For fields smaller than $D_c$ the classical anisotropic dispersion of phosphorene systems is observed with slightly different effective masses, whereas the formation of Dirac-like spectra is observed for fields above the critical value, which is a consequence of the inversion of the conduction and valence bands. The possibility of this transition in phosphorene systems may allow its use as a platform for studying topological phases in 2D systems and for investigating some exotic phenomena, such as unusual Landau levels\cite{ref19, ref20}.  

Although the application of gate voltages in multilayer black phosphorus systems was investigated in recent works\cite{ref11, ref12, ref13, LLL, ref15, ref16}, only a few of them explore the effects of the 
formation of Dirac-like spectra for displacement fields at the vicinity of $D_c$\cite{ref12, ref13} associated with a critical gate density $n_c$. To the best of our knowledge, there is a lack of information concerning these effects in confined nanostructures, such as phosphorene nanoribbons (PNRs), and the role played by them in their electronic and transport properties. Therefore, in this letter, we study the electronic properties of gated multilayer PNRs. In particular, we give special attention to the phase transition occurring at the vicinity of $n_c$ for zigzag and armchair PNRs. We show that the carrier concentration undergoes a change in its behavior for a given transition gate density $n_i$ (for a given Fermi level $E_i$) and investigate the width and thickness dependence of this quantity. Such behavior may be used to determine the number of layers and the orientation of phosphorene based systems. Furthermore, we make a comparison between the results taken from considering screened and unscreened displacement fields and show that the results are qualitatively equivalent.

\section{Theoretical model}\label{sec.model}

\begin{figure}[t]
\centerline{\includegraphics[width = \linewidth]{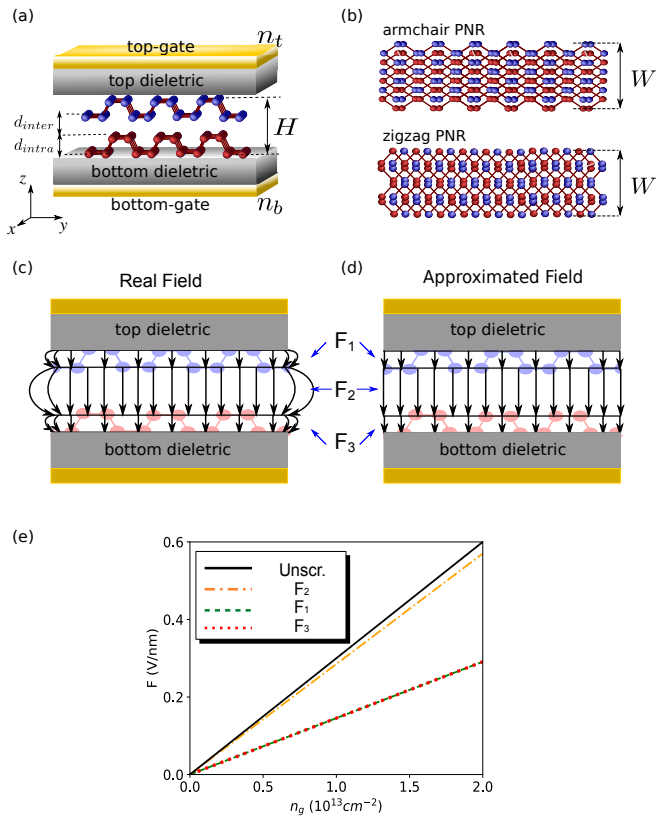}}
\caption{(Color online) (a) Gated bilayer phosphorene nanoribbon. We assume BP nanoribbon systems under the influence of top and bottom gates, with respective gate concentrations $n_t$ and $n_b$, where the intra and interlayer distances are $d_{intra} = 2.153$ \AA\ and $d_{inter} = 3.214$ \AA\ , respectively, and the thickness $H$ is related to the number of layers in the system. (b) top view of armchair (top panel) and zigzag (lower panel) BP nanoribbons with width $W$. A sketch of the real and the approximated screened electric field is presented in (c) and (d), respectively. The density of field lines is related to the field intensity in each region ($F_1$, $F_2$ and $F_3$, as sketched in (c) and (d)). (e) The electric field between each sublayer as a function of the gate density for the screened and unscreened cases from Ref.~[\onlinecite{LLL}]. $F_1$ is represented by the green dashed line, $F_2$ is represented by the orange dash-dotted line, $F_3$ is represented by the red dotted line and the uniform unscreened field is represented by the black solid line.} 
\label{Fig0}
\end{figure}

We consider multilayer PNRs with zigzag (zz) and armchair (ac) edges. The zz (ac) nanoribbon is oriented along the $x$ ($y$) axis, being limited by its width $W$ along the $y$ ($x$) direction and by its thickness $H$ along the $z$ direction, as skecthed in Fig.~\ref{Fig0}(a). The thickness is related to the number of layers $N$ by $H = d_{intra} N + d_{inter}(N - 1)$, where $d_{intra} = 2.153$ \AA\ and $d_{inter} = 3.214$ \AA\ are the intralayer and interlayer distances, respectively.

We employ the tight-binding model proposed by Rudenko \textit{et al.}\cite{ref21} to describe the electronic properties of PNRs in our calculations. Such model shows excellent agreement with DFT-GW calculations for low energy excitations and has been used in several recent works\cite{LLL, ref22, ref23, ref24, ref25}. The Hamiltonian in second quantization assumes the form
\begin{equation}
\mathcal{H} = \sum_{i \neq j} t_{ij} c_i^{\dagger} c_j + \sum_{i\neq j} t_{ij}^{\perp} c_i^{\dagger} c_j + \sum_i U_i c_i^{\dagger}c_i,
\label{Eq1}
\end{equation}
where the operator $c_i$ ($c_i^{\dagger}$) annihilates (creates) an electron in atomic site $i$. Additionally, $t_{ij}$ and $t_{ij}^{\perp}$, are the intralayer and interlayer hopping parameters between atomic sites $i$ and $j$, respectively, given in Refs.~[\onlinecite{ref21, ref25}]. We simulate a perpendicular unscreened electric field by considering the differences in the onsite energies for sublattices located at different heights $z_i$ through the linear formula: $U_{i}= U(z_i) = e(D/\kappa \epsilon_0)(z_i - H/2)$, where $e$ is the fundamental electric charge, $\kappa$ is the dielectric constant of the system, $\epsilon_0$ is the vacuum permittivity and $D$ is the magnitude of the displacement field inside the ribbons. 
Recent studies have shown the effects of the charge redistribution due to voltage gates in multilayer BP\cite{LLL} and show that the screening tends to be more pronounced for systems with larger thicknesses requiring higher critical gate densities, defined as the value for which a gap closure is achieved. In order to incorporate these effects, we make an approximation by considering the screened electric fields calculated for an infinite system and, thus we do not consider the effects of the distortion of the field lines at the edges of the PNRs [See Figs.~\ref{Fig0}(c)]. Figure~\ref{Fig0}(e) shows the electric field between each sublayer of bilayer BP as a function of the gate density. $F_1$ and $F_3$ are the intralayer electric fields, whereas $F_2$ is the interlayer field. As one can notice, the field intensities are different in each region for a given gate density due to the screening effects, where $F_1 = F_3$, as seen from the red dotted and the green dashed curves. The top and bottom gate densities are chosen such that $n_g = n_b = -n_t$, which results in the linear dependence of the electric field on the gate density, $F_i = \beta_i n_g$, where $\beta_i$ is the slope of the field in the i-th region \cite{LLL}. The uniform unscreened electric field dependence is represented by the black solid line. The onsite energies are calculated as:
\begin{equation}
U_i = \sum_{j=1}^{i-1} \beta_{j} d_{j} n_g,
\end{equation} 
where $\beta_1 = \beta_3 = 1.46 \times 10^{-14}$ V$\cdot$cm$^2$/nm and $\beta_2 = 2.85 \times 10^{-14}$ V$\cdot$cm$^2$/nm and $d_1 = d_3 = d_{intra}$ and $d_2 = d_{inter}$ and $i = 1, 2, 3, 4$ is the sublayer index. We also symmetrize the onsite energies by choosing $U_1 = -U_4$.

Even though the screening effects might be different for confined systems, as sketched in Figs.~\ref{Fig0}(c) and (d), we expect our results to be reasonably accurate since we are considering wide PNRs in which edge states are absent. It is worth mentioning that we do not employ the usual procedure of performing self-consistent calculations in order to obtain the onsite potential energies. Instead, we use the explicit gate density-dependent screened displacement fields between each sublayer of the system, as presented in Ref.~\onlinecite{LLL}, in order to compute the $U_i$. All tight-binding calculations were performed by using the KWANT Python package\cite{kwant}.

\section{Results and discussions}

\begin{figure*}[t]
\centerline{\includegraphics[scale = 0.4]{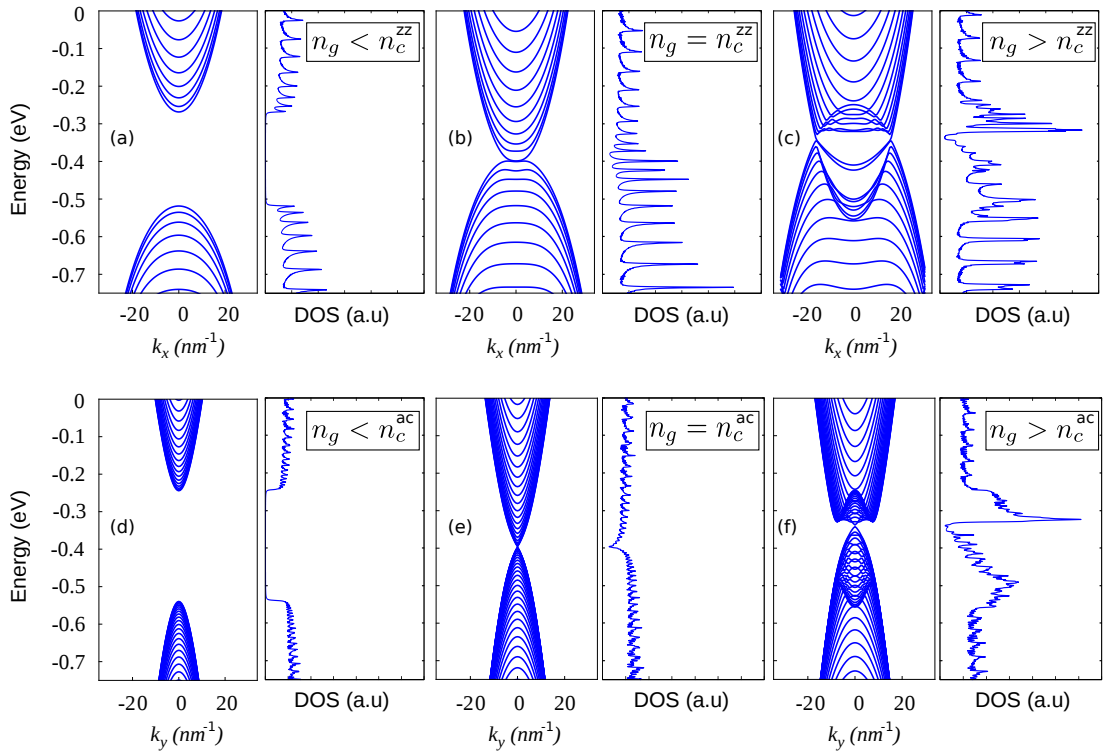}}
\caption{(Color online) Dispersion relation and the correspondent density of states for (a)-(c) zigzag and (d)-(f) armchair phosphorene nanoribbons with various values of $n_g$, for the unscreened case. We consider three different situations: (a) and (d) $n_g < n_c^{zz}$ ($n_g < n_c^{ac}$), (b) and (e) $n_g = n_c^{zz}$ ($n_g = n_c^{ac}$), (c) and (f) $n_g > n_c^{zz}$ ($n_g > n_c^{ac}$), where $n_c^{zz}$ ($n_c^{ac}$) is the critical gate density for which a gap closure is achieved.} 
\label{Fig1}
\end{figure*}

We start by investigating the behavior of the energy dispersion of phosphorene nanoribbons under the influence of gate voltages. We consider the case of bilayer BP nanoribbons in our calculations. Initially, we consider ribbons with fixed width $W = 20$ nm and assume that the charge concentrations on the top and bottom gates ($n_t$ and $n_b$, respectively) have equal absolute values, but with opposite signs $n_t = - n_b = n_g$. It is worth mentioning that the same gate density $n_g$ will generate different fields for the unscreened and screened cases. In the former, $n_g$ will generate an uniform electric field $F = e|n_g|/\kappa\epsilon_0$ between every sublayer of the system, whereas in the latter case, the electric field is constant between adjacent sublayers, not necessarily assuming the same intensity in the different regions\cite{LLL}, as discussed in the previous section. Therefore, we will use the gate density as a means to specify the different regimes instead of the field intensities. In our calculations, we assume $\kappa = 6$\cite{LLL}.

We begin our analysis by considering the unscreened case (solid black curve in Fig.~\ref{Fig0}(e)). Figure~\ref{Fig1} shows the energy bands and the respective density of states (DOS) for zigzag (a)-(c) and armchair (d)-(f) BP nanoribbons with different values of $n_g$. The critical density for zz and ac nanoribbons, $n_c^{zz} = 11.37\times 10^{13}$ cm$^{-2}$ and $n_c^{ac} = 11.17\times 10^{13}$ cm$^{-2}$, respectively, generates critical displacement fields, $D_c^{zz}$ and $D_c^{ac}$, for which a gap closure is achieved. For densities smaller than $n_c^{zz(ac)}$ one observes the typical anisotropic parabolic energy dispersion of phosphorene systems, which leads to nanoribbons with different band curvatures, as one can see in Figs.~\ref{Fig1}(a) and (d). Additionally, the DOS also presents the tipical behavior of gapped systems, showing more pronounced peaks for the zz case than for the ac case. This is due to the smaller band curvature in the zz case, which results in a wider range of momenta in which the Van Hove singularity condition is achieved. At the critical density, a gap closure is observed as one can see in Figs.~\ref{Fig1}(b) and (e) for zz and ac bilayer PNRs, respectively. One can notice that the conduction and valence bands touch  without any significant distortion of the parabolic behavior of the dispersion.

For densities greater than $n_g^{zz(ac)}$, the valence and conduction bands undergo a phase transition resulting in the formation of Dirac-like spectra. Such behavior can be seen in Fig.~\ref{Fig1}(c) and (f) for the zz and the ac cases, respectively.  For sufficiently low energies in both cases, the dispersion resembles those of graphene nanoribbons, in which one observes the presence of two Dirac cones in the zz case and a single point in which the dispersion is approximately linear in the ac case\cite{ref27, ref28}. Additionally, zz graphene nanoribbons have flat bands corresponding to states localized at the edges of the ribbons. However, in the case of gated zz phosphorene nanoribbons the quasi-flat bands that appear between the two quasi-Dirac cones are not related to states localized at the edges of the ribbon, since we are considering ribbons that do not support edge states (beard zz nanoribbons\cite{ref29,ref30}). Interestingly, the DOS has a complicated structure, showing a great asymmetry between the new conductance and valence bands at low energies. This behavior resembles those of massless Dirac Fermions in graphene\cite{ref27}, reinforcing the linear nature of the dispersion in BP at low energies for densities greater than the critical one. 

\begin{figure}[t]
\centerline{\includegraphics[width = \linewidth]{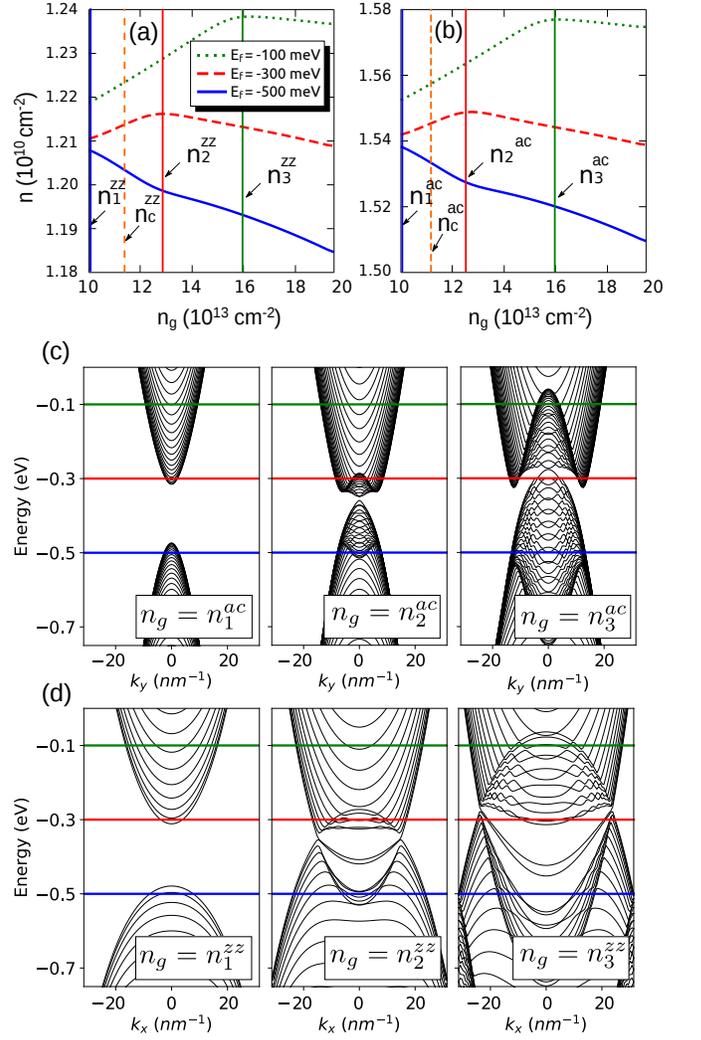}}
\caption{(Color online) Carrier concentration $n$ as a function of the gate density at different Fermi levels for (a) zz and (b) ac bilayer phosphorene nanoribbons. The vertical orange dashed lines mark the critical densties for each case : $n_c^{zz} = 11.37\times 10^{13}$ cm$^{-2}$ and $n_c^{ac} = 11.17\times 10^{13}$ cm$^{-2}$ for zz and ac ribbons, respectively. The other vertical lines mark the transition gate densities. Thus, $n_1^{zz(ac)}$, $n_2^{zz(ac)}$ and $n_3^{zz(ac)}$ are the mentioned densities for the Fermi levels $E_F = -500$ meV, $E_F = -300$ meV and $E_F = -100$ meV, respectively. Figures~(c) and (d) show the band structure for the zz and ac cases, respectively, considering the transition gate densities mentioned above. The horizontal lines in these figures mark the Fermi levels considered in this analysis.} 
\label{Carrier_Concentration}
\end{figure}

The great variation of the DOS at the transition $n_g < n_c \rightarrow n_g > n_c$ suggests that one can achieve significant modulation of the carrier concentration, at a low Fermi level, with a small variation in the gate carrier density. Additionally, one migth expect that the carrier concentration would present a peculiar behavior due to the phase transition at $n_c^{zz(ac)}$. In fact, Figs.~\ref{Carrier_Concentration}(a) and (b) show the dependence of this quantity as function of the gate density for zz and ac bilayer phosphorene nanoribbons, respectively, considering different Fermi levels $E_F = E_i$ (with $E_1 = -500$ meV, $E_2 = -300$ meV and $E_3 = -100$ meV), at $T = 300$ K. In each figure, the orange vertical dashed line marks the position of the critical gate densities. Considering $E_F = E_3$ and $E_F = E_2$, one notices an increase in the carrier concentration even for gate densities greater than $n_c^{zz(ac)}$, as one can see from the dotted green and dashed red curves in both figures. For $E_F = E_1$, one observes a decrease in the carrier concentration of the ribbons. This is due to the fact that the Fermi level is at the valence band region in the latter case, leading to an excess of positive charge carriers in the system (\textit{p}-doped bilayer PNRs). For the \textit{n}-doped bilayer PNRs, the Fermi levels ($E_F = E_2$, $E_F = E_3$) are in the conduction band region, corresponding to an excess of negative charge carriers. Therefore, it is natural to expect that the carrier concentration would increase (decrease) with the gate density for \textit{n}-doped (\textit{p}-doped) systems. Interestingly, this behavior does not hold true for even higher values of $n_g$. In fact, Figs.~\ref{Carrier_Concentration}(a) and (b) show that the excess carrier concentrations for the \textit{n}-doped cases start to decrease for certain gate densities, $n_i^{zz(ac)}$ with $i = 2,3$, beyond $n_c^{zz(ac)}$, which depends on the Fermi level. The gate densities for which the transition occurs are $n_2^{zz} = 12.50 \times 10^{13}$ cm$^{-2}$ ($n_2^{ac} = 12.75 \times 10^{13}$ cm$^{-2}$) and $n_3^{zz} = 16.25 \times 10^{13}$ cm$^{-2}$ ($n_3^{ac} = 16.00 \times 10^{13}$ cm$^{-2}$) for the zigzag (armchair) PNRs. Such behavior is a unique characteristic of the phase transition, which allows for a change in the sign of the curvature of a band for a range of momenta at the vicinity of the $\Gamma$ point. At the conduction (valence) band, a subband with negative (positive) effective mass contributes as hole (electron) bands. Therefore, an excess of positive (negative) charge carriers is added to the negative (positive) excess carrier density of a \textit{n}-doped (\textit{p}-doped) system. For a given Fermi level in the \textit{n}-doped case, the carrier concentration will stop to increase even further whenever the top of the hole subbands touches the Fermi level. Consequently, as the contribution of the holes bands becomes more pronounced than the electron bands, the carrier concentration will start to decrease. In fact, Figs.~\ref{Carrier_Concentration}(c) and (d) show the energy dispersion for the ac and zz nanoribbons, respectively, considering the corresponding gate densities for which the carrier concentration starts to decrease for a given Fermi level. The horizontal lines mark the positions of the assumed Fermi levels : $E_F = E_1$ (blue curve), $E_2$ (red curve) and $E_F = E_3$ (green curve). As one can notice in the case of the \textit{n}-doped BP nanoribbons, for densities greater than $n_c^{zz(ac)}$, there are several energy subbands at the conduction band in which the curvatures are negative. More specifically, for $n_g = n_2^{zz(ac)}$ these subbands start to touch the Fermi level $E_F = E_2$. A similar behavior is observed for the transition gate density $n_3^{zz(ac)}$ and the Fermi level $E_F = E_3$, which is consistent with the non-monotonic behavior of the carrier concentration.

\begin{figure}[t]
\centerline{\includegraphics[width = \linewidth]{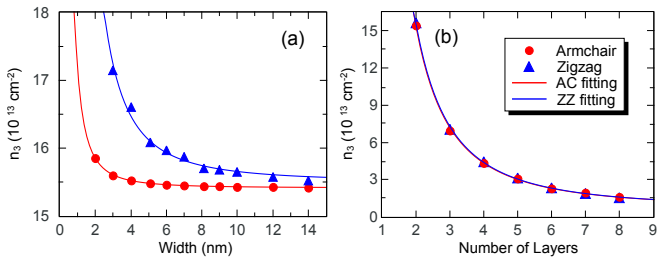}}
\caption{(Color online) (a) Width and (b) thickness dependence of the gate density $n_3^{zz(ac)}$ for zigzag (blue results) and armchair (red results) bilayer BP nanoribbons. The symbols are the results obtained from the tight-binding model, whereas the solid curves are the correspondent fittings.} 
\label{Fig4}
\end{figure}

The properties discussed so far also hold for phosphorene nanoribbons with arbitrary widths and thicknesses. In fact, one can always find a transition gate density $n_{i}^{zz(ac)}$, for a given Fermi level, for zz and ac PNRs for several different sizes. Figure~\ref{Fig4}(a) shows the behavior of the gate density $n_3^{zz(ac)}$ ($E_F = E_3$) for bilayer PNRs with several widths. The symbols are the tight-binding results, where the red circles (blue triangles) correspond to the ac (zz) case, and the solid curves are the fittings. As seen, there is a drop in the value of the transition gate densities with $W$ and despite the difference in the values of $n_3$ for zz and ac cases, the scaling behavior is the same for both cases. The best fit shows a $\propto W^{-2}$ behavior that is quite distinct from what one would expect, since other properties, such as the energy gap of PNRs, present different scaling laws due to the differences in the momentum dependency of the energy levels in the $x$ and $y$ directions\cite{ref29}. A similar picture holds true for the dependence of $n_3^{zz(ac)}$ with the thickness of the ribbons, i.e. with the number of layers $N$. In figure~\ref{Fig4}(b), we show the transition gate density for the chosen Fermi level as a function of the number of layers for a fixed width $W = 20$ nm. For such wide ribbons, the $n_3$ values for zz and ac ribbons are very close to each other. The result shows a clear decrease in $n_3$ for both cases, exhibiting a $N^{-2}$ behavior as in the case of the width dependence. 

The present results suggest a scheme by which one may assess width and the thickness of phosphorene systems: The specific values of the transition gate densities $n_i^{zz(ac)}$ can be easily detected experimentaly through Hall measurements at low magnetic fields. As shown, a small change in $n_g$ beyond $n_i^{zz(ac)}$ causes the carrier concentration to decrease for \textit{n}-doped BP nanoribbons. In this way, the obtained result can be compared with the values of Fig.~\ref{Fig4}, determining the corresponding width and thickness. We expect such method to be accurate for ribbons with small dimensions, since the values of $n_i^{zz(ac)}$ are very close for larger systems. 

It is important to emphasize that the previous results are based on the assumption of an unscreened electric field. In a realistic setting, the charge distribution inside multilayer BP systems generate an electric field that counteracts the previous one, resulting in a screened field. Such an effect plays an important role in the determination of the specific values of the gate critical densities\cite{LLL}. Thus, we proceed to study the differences between the results obtained from unscreened and screened electric fields. As mentioned previously, we approximate the screened electric field inside the nanoribbons by assuming that it is approximately equal to the one of an infinite system. This means that we do not consider the effects of the distortion of the field lines in the presence of the edges [See Figs.~\ref{Fig0}(c) and (d)]. Since we are interested only on the bulk properties of large BP nanoribbons ($d_{inter} \ll W$), we expect the field lines in each region to be very similar to the infinite case. Therefore, we use the screened electric fields from Ref.~[\onlinecite{LLL}] in the case of bilayer BP. 
\begin{figure}[t]
\centerline{\includegraphics[width = \linewidth]{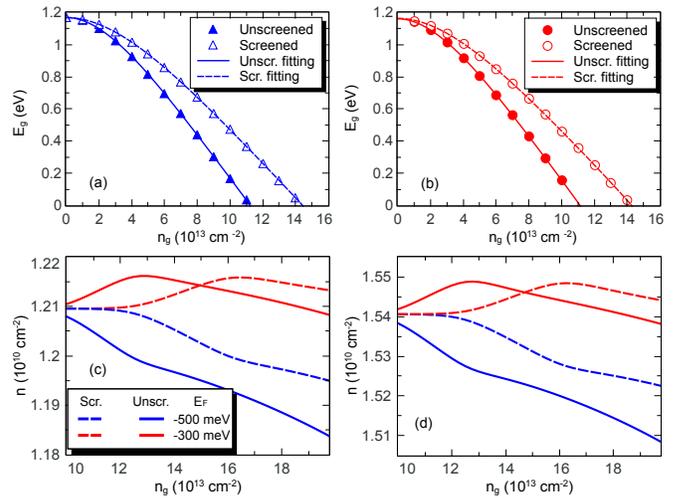}}
\caption{(Color online) Gap as a function of the gate density $n_g$ for (a) zz and (b) ac BP nanoribbons. The closed (open) symbols are the results from the tight-binding model considering the unscreened (screneed) electric field, whereas the solid (dashed) curves are the fittings. The carrier concentration as a function of the gate density is show in (c) for zz, and in (d) for ac nanoribbons. The screened (unscreened) results are represented by the dashed (solids) curves. We considered $E_F = -500$ meV (blue curves) and $E_F = -300$ meV (red curves). } 
\label{Fig5}
\end{figure}

Figures~\ref{Fig5}(a) and (b) show a comparison between the gap dependence of zz and ac PNRs, respectively, for the unscreened and screened fields. The symbols are the results from the tight-binding model and the curves are the corresponding fittings. As one can see, the gap closure is achieved for a higher critical gate density for the screened case as compared with the unscreened case for both, zz and ac BP nanoribbons. This is due to the fact that the fields produced by the charge redistributions within each sublayer counteract the gate field, demanding higher gate densities for a gap closure. The screened critical gate densities are $n_{c, \mathrm{Screened}}^{zz} = 14.43 \times 10^{13}$ cm$^{-2}$ and $n_{c,\mathrm{Screened}}^{ac} = 14.30 \times 10^{13}$ cm$^{-2}$ for zz and ac PNRs, respectively. We also compare the screened and unscreened carrier concentrations as a function of the gate densities for \textit{n}-doped ($E_F = E_2$) and $p$-doped ($E_F = E_1$) bilayer BP nanoribbons. Figures~\ref{Fig5}(c) and (d) show the results for the zz and ac cases. The screened results, represented by the dashed lines, present the same behavior as the unscreened results (solid lines), i.e. there is a gate density $n_{i,\mathrm{Screened}}^{zz(ac)}$ for which the carrier concentration start to change its behavior. The difference is that $n_{i,\mathrm{Screened}}^{zz(ac)} > n_{i}^{zz(ac)}$, for a given Fermi level $i$, as one should expect. Therefore, we can conclude that the screened results are qualitatively equivalent to the unscreened results. 

\section{Conclusions}

We have investigated the electronic properties of gated BP nanoribbons with zz and ac edges. We have shown that for gate densities beyond a critical value $n_{c}^{zz(ac)}$ the energy bands undergo a phase transition resulting in the formation of a Dirac-like spectrum. Such transition is shown to strongly affect the electronic properties of these systems leading to an anomalous behavior of the gate-induced excess carrier concentration. For instance, the carrier concentration of a gated \textit{n}-doped BP nanoribbon increases with the gate density until certain value $n_i^{zz(ac)}$, which depends on the Fermi level $i$, and start to decrease for higher gate charge concentrations. We have also studied the scaling laws of $n_i^{zz(ac)}$ with the width $W$ and the number of layer $N$ for ribbons with zz and ac edges. Our results show a $W^{-2}$ and a $N^{-2}$ dependencie for nanoribbons with both types of edges. We discuss how the combination of such results could possibly be used to determine the width and thickness of BP nanoribbons. 
\section{acknowledgements}

This work was financially suported by the CAPES foundation. The authors thank Dr. L. L. Li for the valuable discussions.

\end{document}